\documentstyle[12pt,twoside,fleqn,espcrc2]{article}
\title{Nucleosynthesis and Gamma Ray-Line Astronomy}
\author{Elisabeth Vangioni-Flam\address{
Institut d'Astrophysique de Paris,\\
98bis Bd Arago 75014 Paris, France}
, Reuven Ramaty\address{GSFC, NASA, Greenbelt, MD 20771, USA}
and
Michel Cass\'e\address{
Service d'Astrophysique, CEA,\\
Orme des Merisiers 91191 Gif/Yvette, France}}

\begin{document}

\maketitle

The most energetic part of the electromagnetic spectrum bears the 
purest clues to the synthesis of atomic nuclei in the universe. The 
decay of radioactive species, synthesized in stellar environments 
and ejected into the interstellar medium, gives rise to specific 
gamma ray lines. The observations gathered up to now show evidence 
for radioactivities throughout the galactic disk, in young supernova 
remnants (Cas A, Vela), and in nearby extragalactic supernovae (SN 
1987A, SN 1991T and SN1998bu), in the form of specific gamma ray 
lines resulting, respectively, from the radioactive decay of 
$^{26}$Al, $^{44}$Ti and $^{56}$Co. The various astrophysical sites 
of thermal nucleosynthesis of the radioactive nuclei were discussed: 
AGB and Wolf-Rayet stars, novae, and type Ia and type II supernovae. 
Nuclear excitations by fast particles also produce gamma ray lines 
which have been observed in great detail from solar flares, and more 
hypothetically from active star forming regions where massive 
supernovae and WR stars abound. This non thermal process and its 
nucleosynthetic consequences was reviewed. The 511 keV line arising 
from e$^+$ + e$^-$ annihilation also provides important information 
on explosive nucleosynthesis, as well as on the nature of the 
interstellar medium where the positrons annihilate. INTEGRAL, the 
main mission devoted to high resolution nuclear spectroscopy, should 
lead to important progress in this field.


\section{Observational status}

The experimental situation in gamma ray line astronomy was 
summarized by G. Vedrenne. The highlights are: i) the discovery (1) 
of  $^{44}$Ti emission from the Vela region (GRO JO852-4642) near a 
new supernova remnant detected in X rays by the ROSAT satellite (2); 
ii) the positive detection of a recent SNIa by COMPTEL (SN 1998bu) 
located at 8.1 Mpc (3); iii) the release of a new $^{26}$Al COMPTEL 
map derived from the observations using a sophisticated technique of 
data analysis (4); iv) the withdrawal of the Orion gamma ray line 
data, followed immediately by the announcement of a similar emission 
from the Vela region (5). In addition, P. von Balmoos presented a 
review on the origin of galactic positrons, including compact 
galactic sources and radioactive nuclei ($^{26}$Al, $^{44}$Ti, 
$^{56}$Co).

\section{Production of radioactive nuclei in thermal nucleosynthesis}

\subsection{Non explosive nucleosynthesis: AGB and Wolf-Rayet stars}

G. Meynet analyzed the synthesis of $^{26}$Al in AGB and WR 
stars. Production in AGB stars falls short from explaining the 
required live radioactive aluminum in the galaxy (about 2M$_\odot$), 
but WR stars remain a serious candidate. Indeed, the detailed 
analysis of the COMPTEL $^{26}$Al map and its correlation with the 
free-free emission of the galactic disk, as observed by COBE,  
indicates that massive stars are the most likely candidates for 
$^{26}$Al production (4). But at the moment, it is not possible to 
discriminate between core collapse supernovae and WR stars since 
neither the Vela SNR nor the $\gamma$ Velorum WR star coincide with 
peaks on the 1.8 MeV COMPTEL map. The absence of a clear detection 
signal implies that the progenitor of WR11 in Vela has a mass less 
than 40 M$_\odot$.

M. Arnould, broadening the scope, has pointed out the exceptional 
interest of radionuclide astrophysics at large, since it provides 
strong links between gamma ray astronomy, chemical evolution of the 
galaxy, stellar nucleosynthesis, and the physico-chemistry of 
circumstellar envelopes, the ISM, and the early solar system. After 
a critical analysis of all nucleosynthetic sites, he concluded that 
WR modeling is immensely simpler than that of AGB stars, novae and 
supernovae. He surmised that WR stars might be of interest to 
cosmochemists since they could provide a wealth of isotopic 
anomalies, potentially observable in meteorites.

\subsection{Explosive nucleosynthesis}
 
A. Core collapse supernovae and their remnants (SNII).

The nucleosynthesis of $^{26}$Al and $^{44}$Ti by core collapse was 
critically examined by F-K. Thielemann. Taking for example a 15 
M$_\odot$ star, the $^{26}$Al yields of (6) and (7) differ 
significantly (3$\times$10$^{-8}$ against 2.7$\times$10$^{-6}$ 
M$_\odot$). The origins of differences concern  the choice of the 
still controversial $^{12}$C($\alpha,\gamma$)$^{16}$O  reaction 
rate, and above all the treatment of convection (Schwarzschild or 
Ledoux + semiconvection, rotationally induced mixing and so on). 
Concerning the Fe-group elements, the variations between models are 
expected to be more acute due to a different simulation of the 
explosion, affecting the mass cut. Surprisingly, for the 15 
M$_\odot$ model, the amounts of ejected $^{44}$Ti, $^{56}$Ni, 
$^{57}$Ni are similar in both cases (respectively 
6$\times$10$^{-5}$, 0.1 and 4$\times$10$^{-3}$ M$_\odot$). However, 
the optimism should be tempered since the $^{44}$Ti yield varies a 
lot as a function of mass between the different authors. Anyway, 
using the new half-life determination (59-62 yr), the amounts of 
$^{44}$Ti ejected by SN 1987A (estimated from the late light curve, 
roughly 10$^{- 4}$ M$_\odot$), Cas A (about 1.3$\times$10$^{-4}$ 
M$_\odot$ from gamma rays) and JO852-4642 in Vela 
(5$\times$10$^{-5}$ M$_\odot$) can be explained with a calculation 
employing spherical symmetry. Concerning the synthesis of Fe, the 
main question is whether the mass of $^{56}$Ni ($^{56}$Fe) ejected 
by core collapse supernova decreases or not as a function of the 
mass of the progenitor above 20 M$_\odot$.  Light curve analyses 
that should help to solve this question are for the moment limited 
to the low mass range (less than 30 M$_\odot$), unfortunately. So 
the question remains unsettled.

B. Thermonuclear supernovae (SNIa).

Type Ia supernovae are expected to produce greater quantities of 
$^{56}$Ni and to become transparent to gamma rays earlier than their 
gravitational counterparts of high masses (SNII, SNIb, SNIc), as was 
discussed by S. Kumagai. Thus they are good targets of opportunity 
for gamma ray line astronomy. Indeed, an unusually bright SNIa 
(SN1991T), located at 13 - 17 Mpc, was already observed by 
COMPTEL at the edge of the Virgo cluster, close to the detection 
limit. The amount of ejected $^{56}$Ni derived from the observation 
(higher than 1 M$_\odot$) appears quite unusual, as does SN1991T 
itself. SN 1998bu, at a distance of 8.2 Mpc, presents certain 
similarities to SN 1991T. However, contrary to SN 1991T, the mean 
847 keV line flux observed by COMPTEL from 5 to 131 days after 
the appearance of the supernova is somewhat low compared to the 
predictions of the models.

Concerning the observability of type Ia supernovae by INTEGRAL, S. 
Kumagai gave a mildly optimistic view. However, recent work (8), 
that took into account the width of the 847 keV $^{56}$Co decay 
line, tempers this enthusiasm somewhat. If, by chance, an SNIa is 
captured by INTEGRAL in good conditions, the observation would help 
to calibrate the explosion models, the ejecta structure and the 
$^{56}$Ni distribution.
                                                                               
C. Novae

The best prospects are the lines resulting from $^7$Be and $^{22}$Na 
decays. None of these have been observed up to now. Only upper 
limits exist (2$\times$10$^{-8}$ M$_\odot$) on the ejected $^{22}$Na 
from neon rich novae. The GRO sky survey at 1.275 MeV gives only a 
marginal excess from South Aquila.  M. Hernanz presented detailed 
nucleosynthesis calculations in nova explosions, employing 
hydrodynamical models for a variety of CO and ONe white dwarf 
masses. The low ejected mass of $^{22}$Na obtained in the ONe model 
is consistent with the observational upper limit. Only nearby novae 
should be captured by INTEGRAL, through the radioactive decay of 
$^7$Be (CO novae: 500 pc) and $^{22}$Ne (1.5 kpc, (9)). In all 
models a strong continuum dominates the gamma ray spectrum during 
the early period of expansion. This short and intense emission could 
be detected at least up to 3 kpc, during a few hours (10).

\section{Non thermal gamma ray lines and associated nucleosynthesis}

A broad overview of all aspects of gamma ray lines induced by non 
thermal particles in various astrophysical sites, including solar 
flares, was presented by R. Ramaty. The $^{12}$C and $^{16}$O lines, 
at 4.438 and 6.129 MeV, can only be produced by non thermal particle 
interactions, a fact that can be used to distinguish a nonthermal 
from a nucleosynthetic origin of an observed gamma ray line 
spectrum. The $^{12}$C and $^{16}$O lines, as well as many others 
have been observed from solar flares. The most prominent ones are at 
2.223 MeV following neutron capture on H, at 0.511 MeV from positron 
annihilation (both the neutrons and positrons results from 
nonthermal particle interactions), at 1.634 MeV from $^{20}$Ne and 
at 0.429 and 0.478 MeV from $^7$Be and $^7$Li produced in 
interactions of fast $\alpha$ particles with He. These lines have 
provided much new information on particle acceleration as well as on 
the properties of the solar atmosphere (11, 12). With the withdrawal 
of the COMPTEL observations of the $^{12}$C and $^{16}$O lines from 
Orion, there remains no convincing evidence for such lines from 
non-solar sites. But the fast particles which produce the lines 
could have an important role in the origin of some of the light 
elements, in particular Be (13). This was the subject of a recent 
conference, the proceedings of which should appear shortly (see 14). 

\section{Conclusion}

The COMPTON GRO mission has provided a wealth of data which has 
given a strong impetus to nuclear astrophysics. Now a new episode is 
opening up with INTEGRAL.  In this context,  F. Lebrun has shown the 
potential of the INTEGRAL satellite. The high quality spectroscopy 
of the SPI instrument, between 2 keV and 1 MeV, will shed light on 
fundamental questions of nucleosyntheis. Lines from $^{44}$Ti decay 
will be observed with both the SPI spectrometer and the IBIS imager.
The proceedings of the invited talks and posters are available in the CDROM 
of the Texas Symposium.


\begin{thebibliography}{9}
\bibitem{iyu} Iyudin, A.F. et al. 1998, Nature, 396, 142.
\bibitem{asc} Aschenbach, B. 1998, Nature, 396, 141.
\bibitem{lei} Leising, M.D. 1998 Third INTEGRAL Symposium, "The
extreme Universe", Taormina, to be published.
\bibitem{knod} Knodlseder, J. 1997, Thesis, Toulouse University.
\bibitem{meu} van der Meulen, R.P. et al. 1998, Third INTEGRAL Symposium,
"The Extreme Universe", Taormina, to be published.
\bibitem{woo} Woosley, S.E. and Weaver, T.A. 1995, ApJS, 101, 181.
\bibitem{Thi} Thielemann, F.K. et al. 1996, ApJ, 460, 108.
\bibitem{ise} Isern, J., 1998, Third INTEGRAL Symposium, "The Extreme
Universe", Taormina, to be published.
\bibitem{jos} Jos\'e, J. and Hernanz, M. 1998, ApJ, 494, 680.
\bibitem{gom} Gomez-Gomar J. et al. 1998, MNRAS, 296, 913.
\bibitem{ram} Ramaty, R. et al. 1995, ApJ, 455, L193. 
\bibitem{Man} Mandzhavidze, N. et al. 1997, ApJ, 489, L99.
\bibitem{evf} Vangioni-Flam et al. 1998, AA, 337, 714
\bibitem{ram2} Ramaty, R. et al. 1999, Publ. Astron. Soc. Pacific, in
press). 
\end{thebibliography}
\end{document}